\documentclass[5p]{elsarticle}
\makeatletter
\def\ps@pprintTitle{%
 \let\@oddhead\@empty
 \let\@evenhead\@empty
 \def\@oddfoot{}%
 \let\@evenfoot\@oddfoot}
\makeatother

\usepackage{csquotes}
\usepackage{graphicx,color}
\usepackage{float}
\usepackage[caption=false]{subfig}
\usepackage{amsmath}
\usepackage{amssymb}
\usepackage{multirow}
\usepackage{dsfont}
\usepackage{adjustbox}
\usepackage{pdflscape}
\usepackage{wasysym}
\usepackage{textcomp}
\usepackage{enumitem}   
\usepackage[percent]{overpic}
\begin{document}

\title{Modeling of runaway electron induced damage on boron-nitride tiles in WEST}
\author{T. Rizzi$^a$, S. Ratynskaia$^a$, P. Tolias$^a$, Y. Corre$^b$, M. Diez$^b$, M. Firdaouss$^b$, J. Gerardin$^b$, R. Mitteau$^b$, C. Reux$^b$, A. Kulachenko$^c$, the WEST team$^d$ ,the EUROfusion Tokamak Exploitation Team$^e$}
\address{$^a$Space and Plasma Physics - KTH Royal Institute of Technology, Teknikringen 31, 10044 Stockholm, Sweden\\ 
$^b$CEA, IRFM, St-Paul-Lez-Durance, France \\
$^c$Mechanics - KTH Royal Institute of Technology, Teknikringen 8, 10044 Stockholm, Sweden\\
$^d$See http://west.cea.fr/WESTteam\\
$^e$See the author list of  E. Joffrin et al 2024 Nucl. Fusion 64 112019 
}

\begin{abstract}
The runaway electron (RE) - induced damage on boron nitride (BN) tiles mounted on the inner bumpers of the WEST tokamak is modeled employing available empirical input and experimental constraints, concerning the post-mortem documentation of the damaged material topology and infra-red camera observations of the long-time decay of the surface temperature.  
A newly developed work-flow for the modeling of brittle failure due to RE impacts, recently validated against a controlled DIII-D experiment, is employed. Monte Carlo simulations of RE transport into BN provide volumetric heat source maps for finite-element simulations of the linear thermoelastic material response, while the brittle failure onset is predicted on the basis of the Rankine criterion. The physics of thermal stress driven failure and explosion are well captured by this model, which exhibits high sensitivity to RE impact parameters. Despite the accidental nature of the damage events, the workflow predicts failure in accordance with observations for realistic loading specifications expected in WEST disruptions.
\end{abstract}
\maketitle

\section{Introduction}

Plasma facing component (PFC) damage inflicted by runaway electron (RE) incidence constitutes a frontier topic in the context of the provision of PFCs with sufficient lifetime, whose understanding is essential for the avoidance of disastrous in-vessel loss-of-coolant accidents\,\cite{Pitts2024,Krieger_2025,Ratynskaia_2025_R}. Modeling of the macroscopic motion of the shallow melt pools created under powerful transient thermal loads has been extensively validated against numerous experiments (see summary in Table 3 of Ref.\cite{Krieger_2025}), lending confidence in the predictive power of the developed tools\,\cite{Ratynskaia2020,Ratynskaia2022_1,Paschalidis2023,Komm_2020}. Meanwhile, due to the extreme intensity of the ultra-localized heat flux and the volumetric nature of the relativistic electron energy deposition, RE impacts lead to explosive PFC damage, and as a consequence, until recently no dedicated controlled experiments were attempted\,\cite{Krieger_2025, Ratynskaia_2025_R} with exception of an early TEXTOR work\,\cite{Forster_2011}. With accidental damage evidence available, modeling of RE-induced PFC damage has been restricted to the thermal response\,\cite{Bartels1994,Cardella2000,Maddaluno2003}. 

The situation recently changed with the first controlled well diagnosed experiment in DIII-D~\cite{Hollmann2024}, where a part of a graphite sample has been blown off by the incident REs and the subsequent release of debris has been observed by infra-red (IR) cameras. The experiment provided several empirical constraints (measured conducted energy, damage topology, explosion time, blown-off material volume, debris speeds) and enabled the validation of a newly developed work-flow for simulations of RE-induced brittle failure\,\cite{Ratynskaia_2025}. For brittle sublimating materials, modeling has already progressed to thermomechanical simulations of the fragmentation process and the debris release\,\cite{ITPA_Rizzi, EPS_Ratynskaia}. The simulations of the thermomechanical response of tungsten (W), a ductile material at high temperatures with a liquid phase, pose several challenges, as outlined in Ref.\cite{Ratynskaia_2025_R}. However, model developments are now expected to benefit from the first controlled RE-induced W damage experiment in WEST (April 2025)\,\cite{EPS_Ratynskaia,ITPA_Corre}. 

In this work, we further test the work-flow capability to describe RE-induced damage in brittle sublimating PFCs by modeling accidental RE impact events on boron nitride (BN) tiles which were observed on the inner limiter of the WEST tokamak and led to significant material loss as well as the release of fast debris. With realistic input for the RE parameters and the PFC wetting, the simulations confirm the realization of thermal-stress driven failure, as proposed earlier based on accidental explosive events in FTU\,\cite{De_Angeli_2023} and controlled explosive events in DIII-D\,\cite{Ratynskaia_2025}. In spite of limitations in the RE impact input due to the accidental nature of these events, the workflow delivers robust predictions consistent with observations for realistic loading specifications expected in WEST disruptions.

\section{Experimental evidence and reconstruction of the RE impact parameters }\label{sec:exp_ev}

\begin{figure}[!h]
    \centering
    \includegraphics[width=0.81\linewidth]{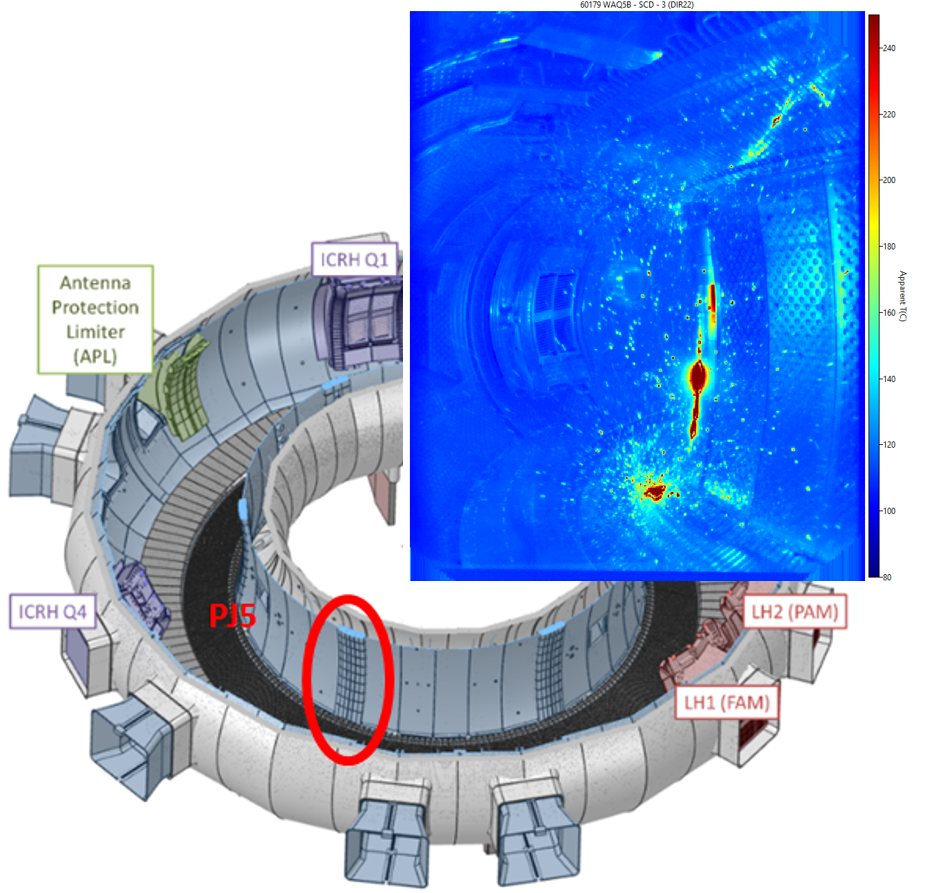}
    \caption{Location of the PJ5 inner bumper in WEST tokamak. IR camera view of the RE beam impacting the limiter and following debris ejection during discharge \#\,60179.}
    \label{fig:tokamak_innerLimiter}
\end{figure}

Here we are concerned with damage on the BN tiles of the inner bumpers of the WEST tokamak inflicted during the runaway experiment executed in 2024 (C9 experimental campaign). During this experiment, several discharges terminated with RE beam losses on the vessel wall, resulting in varying degrees of damage across different locations, both poloidally and toroidally. This study focuses on the inner bumper PJ5, whose toroidal position is indicated in Fig.\ref{fig:tokamak_innerLimiter}. WEST tangential wide angle viewing IR camera\,\cite{Houry_2023} enables observations of RE beam impacts and subsequent ejection of debris, as exemplified in Fig.\ref{fig:tokamak_innerLimiter}.

\begin{figure}[!h]
    \centering
    \begin{overpic}[width=0.70\linewidth]{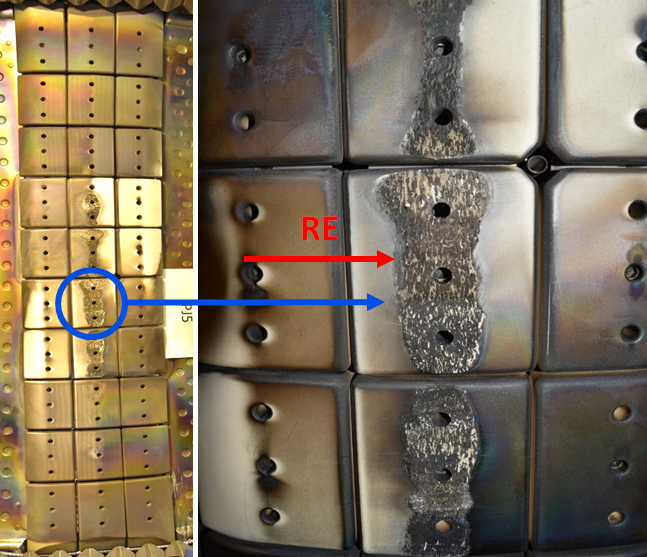}
        \put(20,75){\color{white}\textbf{(a)}} %
        \put(35,75){\color{white}\textbf(b)} 
    \end{overpic}
    \caption{Post-exposure image of the PJ5 limiter with the most damaged tile circled in blue (a), close-up view of the most damaged tile also indicating the direction of the incoming RE beam (b).}
    \label{fig:damage_limiter}
\end{figure}

\begin{figure}[!h]
  \centering
  \begin{minipage}[t]{0.20\textwidth}
    \centering
    \includegraphics[width=\linewidth]{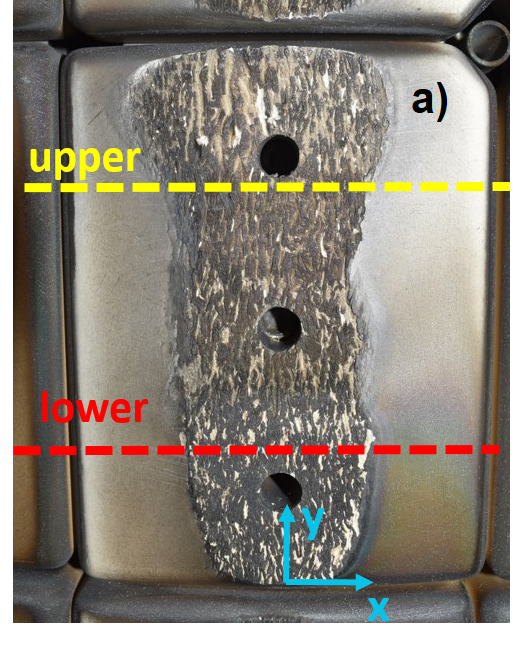}
  \end{minipage}%
  \begin{minipage}[t]{0.3\textwidth}
    \centering
    \includegraphics[width=\linewidth]{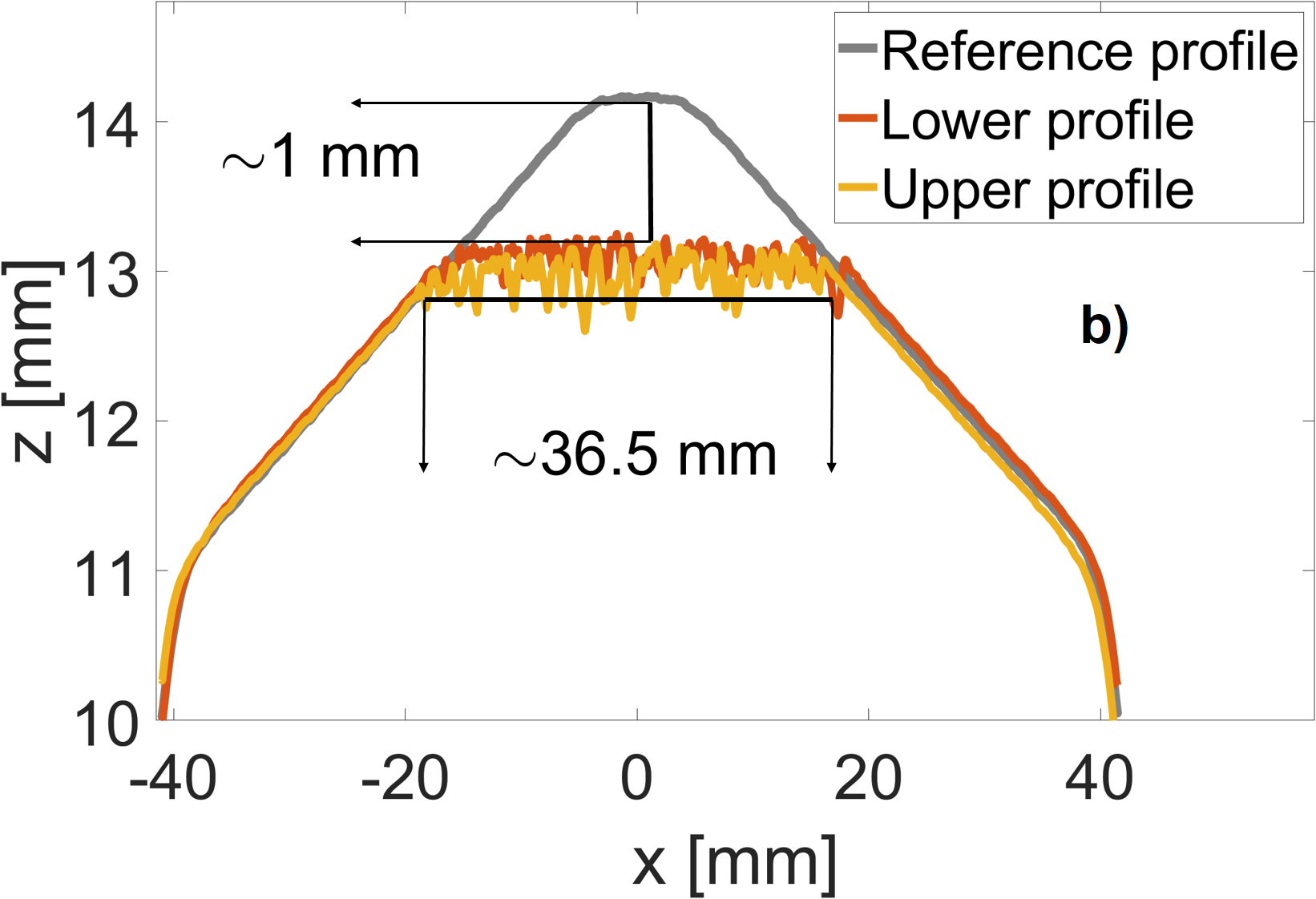}
  \end{minipage}
  \caption{Close-up view of the most damaged tile B6 featuring the lines of the profilometry measurements (a). Results of the profilometry measurements (b). }
  \label{fig:profilometry}
\end{figure}

Post-mortem analysis of the limiter reveals significant erosion across the central column of the tiles, as seen from Fig.\ref{fig:damage_limiter}, with tile B6 being the most severely affected (circled in blue). The observed damage resulted from multiple discharges — at least three — with RE beams terminating on the same tile, as inferred from IR camera observations.
Profilometry measurements on tile B6 reveal symmetric damage relative to the tile’s central axis, with maximum erosion depth of approximately 1 mm, as presented in Fig.\ref{fig:profilometry}. 

\begin{figure}[!h]
    \centering
    \includegraphics[width=0.8\linewidth]{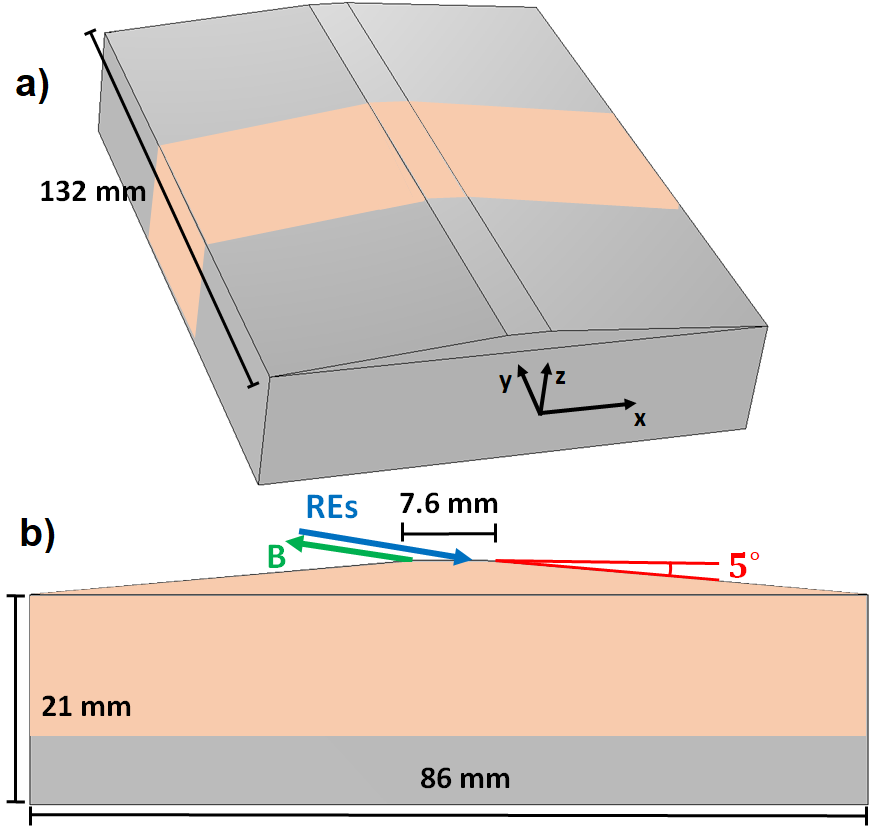}
    \caption{Sketch of the tile geometry; 3D view (a), side view (b). The green arrow indicates  the B-field direction, inclined with an angle of 0.3$^\circ$ with respect to the flat top. The blue arrow indicates the RE beam. The volume in the Geant4-COMSOL simulations, corresponding to the scenario of 20 MeV REs, is highlighted in orange.}
    \label{fig:workflow_geom}
\end{figure}

Magnetic equilibrium reconstruction prior to the disruption indicates a B-field incident angle of 0.3$^\circ$ relative to the flat top of the BN tile, oriented in the direction opposite to the incoming RE beam, see the sketch in Fig.\ref{fig:workflow_geom}. The magnetic field strength at the position of the tile is $4\,$T. The surface area of the tile directly impacted by the REs is not known. However, tracing magnetic field lines up to their intersection with the tile surface allows to exclude wetting of the opposite tile edge.

Since no information is available concerning the RE energy and pitch distributions upon the impact, in this work representative initial kinetic energies are selected and the same approach is adopted for the initial pitch angle with respect to the magnetic field. In particular, we probe the BN thermo-mechanical response to monoenergetic REs of 1 MeV, 10 Mev and 20 MeV with pitch angle in the range $0-25^{\circ}$. This is consistent with initial results of REIS (Runaway Electron Imaging
Spectrometry \cite{Causa_2019}) data analysis \cite{Regine_poster}.  The choice of latter values is motivated by the post-mortem detection of activated W radioisotopes necessitating the presence of high-energy (exceeding $8\,$MeV) REs, also consistent with WEST observations from the wide angle IR camera, see Sec.9 of Ref.\cite{Ratynskaia_2025_R} for further details. 

Due to the accidental nature of the damage event, no data are available to directly quantify the energy deposited by the RE beam on the inner bumper PJ5, and specifically on tile B6. However, the total RE current is inferred to be approximately 200 kA. This can be used to calculate the total number of electrons in the beam and, assuming a mono-energetic distribution, to infer the total kinetic energy carried by the RE population. In addition to this kinetic component, magnetic energy conversion must also be considered, potentially increasing the total available energy by up to a factor of four. This energy is then distributed across multiple RE impact sites. Assuming a uniform distribution across three toroidal limiters, with a few tiles affected at each site, the resulting energy deposition per tile is estimated to range within $2-20\,$kJ for 2 MeV electrons, and $20-200$\,kJ for 20 MeV electrons. Since the high energy fraction of the RE distribution is smaller than the lower energy part, in the simulations $20\,$kJ is chosen as a representative value of the deposited energy. 

Moreover, the signal coming from the neutron detector features a spike with a duration of approximately $\sim$1 ms, which will be assumed as the impact duration of the RE beam on the PJ5 bumper. This is also consistent with the recent WEST experiment on controlled RE-induced damage where an impact duration of 3 ms was inferred from the fast visible camera\,\cite{ITPA_Corre}.

%\begin{figure}[!h]
%    \centering
%    \includegraphics[width=0.99\linewidth]{FIGURES/geometry.png}
%    \caption{Geometry of Boron-Nitride tile. Top view (right) and side view (top).}
%    \label{fig:tile_geom}
%\end{figure}

%\begin{figure}[!h]
%    \centering
%    \includegraphics[width=0.99\linewidth]{FIGURES/PROFILOMETRY.png}
%    \caption{Profilometry measurements of tile B6}
%    \label{fig:profilometry}
%\end{figure}

\section{Modeling workflow} \label{ssec:workflow}

Here we employ the modeling work-flow first reported in Ref.\cite{Ratynskaia_2025}, where it successfully reproduced the controlled RE-induced damage on the graphite sample exposed in the DIII-D tokamak. The first step - modeling of RE orbits by KORC\,\cite{Carbajal_2017,Beidler_2020} is not available for the accidental damage of concern and instead the impact characteristics are reconstructed, as detailed in Sec.\ref{sec:exp_ev}. Thus, the reduced workflow involves two-steps: first, the assessment of the three-dimensional RE energy deposition profiles within the sample; second, the use of the resulting volumetric energy map as input for finite element simulations of the tile thermo-mechanical response. The implementation of the problem in Geant4 and COMSOL is presented in what follows.

\subsection{RE volumetric energy deposition}

Monte Carlo simulations of the volumetric energy deposition are performed with the open-source Geant4 code \cite{Geant4_2003,Geant4_2006,Geant4_2016}. Geant4 enables reconstruction of the full 3D geometry of the tile and the magnetic field topology (both inside and outside the tile volume), as shown in Fig.\ref{fig:workflow_geom}. For simplicity, the three bolts present in each tile are omitted from the model, and fillets are approximated with sharp edges. Due to uniform loading, only a fraction of the tile along the y-direction is simulated, corresponding to about two Larmor radii to ensure the capturing of re-deposited backscattered electrons. Along the z-direction, only the first 14 mm from the top are considered, since essentially no energy is deposited below. In Fig.\ref{fig:workflow_geom}, the simulated domain is highlighted in orange, corresponding to the scenario of a $20\,$MeV RE beam ($R_{\mathrm{L}}\sim17$\,mm). 

The domain is adaptively meshed to capture the sharp near-surface spatial gradients of the energy deposition. The mesh resolution in the most refined regions is determined by the minimum between two characteristic length scales: the thermal diffusion length over the deposition time and the electron range in boron nitride within the continuous slowing down approximation (CSDA) at the relevant energies\,\cite{ESTAR}. As a result, the mesh sizes near the loaded region reach values of $\sim$15\,$\mu$m, gradually increasing up to 500\,$\mu$m in regions further away.

The RE beam is modeled as a uniform particle source wetting a specific surface area, which can be adjusted to represent different impact scenarios. Several cases are explored, with variations in the extent of the tile surface exposed to the RE beam. The opposite side of the tile remains unaffected in all cases due to geometrical shadowing.
Different incident RE kinetic energy and pitch angle combinations have been probed, also accounting for the gyro-phase randomization.

%Experimental observations reveal relatively uniform and symmetric damage along the main axis of the tile. Consequently, the  loading from the runaway electron population is expected to exhibit similar symmetry.
%As a result, it is not necessary to simulate or mesh the entire tile along its main axis. Instead, modeling a region spanning at least two Larmor radii—calculated at the highest simulated kinetic energy and maximum pitch angle—is sufficient to capture any potential asymmetries introduced by the magnetic field. This approximation reduces computational cost and improves Monte Carlo statistics in the region of interest, thereby decreasing statistical noise in the results.

Our Geant4 simulations have been benchmarked against low-Z (such as Be and C) calorimetry measurements with primary electron energy ranging between $0.1-1\,$MeV\,\cite{Lockwood_1980}, as well as against normal and oblique electron backscattering measurements in the keV range\,\cite{Bronshtein_1969,Gomati_2008,Hunger_1979,Neubert_1980}, and against normal electron backscattering experiments in the MeV range\,\cite{Tabata_1967}. As reported for the low-Z graphite\,\cite{Ratynskaia_2025}, different physics libraries (PENELOPE or Livermore) and different scattering implementations (single, mixed, multiple) yield nearly identical energy density maps. Thus, the reported Geant4 simulations employ the Urban scattering model combined with the PENELOPE library\,\cite{Penelope}, to ensure high accuracy and computational efficiency.

\subsection{Thermo-mechanical response}

The coupled thermo-mechanical simulations are carried out with the COMSOL multiphysics software\,\cite{COMSOL_manual}. The 3D energy density maps of Geant4 are converted into power density profiles under the assumption of uniform loading over $1\,$ms. These power maps constitute the volumetric source term in the temperature equation. The heat solver is coupled to the Navier equation for the displacement via thermal expansion; the reader is addressed to Ref.\cite{Ratynskaia_2025} for details on the linear thermoelastic model. 

The temperature dependence of the specific isobaric heat capacity, thermal conductivity and coefficient of thermal expansion for hot-pressed BN is considered, while the mass density is considered to be constant at its room temperature value of $2.0\,$g/cm$^3$. Concerning the specific heat capacity $c_{\mathrm{p}}$, the NIST-JANAF recommendations are followed\,\cite{Chase_1998}. They correlate well with the low temperature measurements of Dworkin \emph{et al.} ($20-300$\,K)\,\cite{Dworkin_1954}, the intermediate temperature measurements of McDonald and Stull ($300-1650\,$K)\,\cite{McDonald_1961}, and the high temperature measurements of Prophet and Stull ($1300-2200$\,K)\,\cite{Prophet_1964}. Note that the NIST-JANAF tabulations assume that the $c_{\mathrm{p}}$ value remains constant above $2200\,$K\,\cite{Chase_1998}. Given this extrapolation, the data are fitted into a half sigmoid function of the form $a(1-be^{-cT})$. Concerning the thermal conductivity $k$, the hot-pressed BN curve within $250-1800\,$K reported by Pierson\,\cite{Pierson_1996} is followed. A quadratic fit is carried out that can be reasonably extrapolated up to 4000 K. It is noted that, for hot-pressed BN, the anisotropy in the thermal conductivity is weak and thus negligible. In particular, $k$ perpendicular to the pressing direction has been reported to be 20-30\% larger than $k$ parallel to the pressing direction\,\cite{Simpson_1976}. Concerning the thermal expansion coefficient $\alpha_{\mathrm{l}}$, hot-pressed BN is characterized by a moderately high thermal expansion parallel to the pressing direction but extremely low thermal expansion perpendicular to it\,\cite{Pease_1952,Kuznetsova_1967}. Thus, the $\alpha_{\mathrm{l}}$ anisotropy is strong and cannot be neglected. Perpendicular to the pressing direction, dedicated dilatometric measurements were carried out within 300-1300K (featuring both negative and positive values), which revealed a linear correlation. Parallel to the pressing direction, a constant value is adopted. It is pointed out that the anisotropy is along the main tile axis, where the loading (and correspondingly the temperature distribution) is uniform, hence no major effects are observed. 

Similarly to the brittle ATJ graphite\,\cite{Ratynskaia_2025}, the constitutive relation for BN is assumed to be linear-elastic until fracture, where the strain is computed also accounting for the thermal contribution. The mechanical properties of BN are necessarily considered to be temperature independent in absence of experimental data above room temperature\,\cite{Pierson_1996,Taylor_1955}. Hot-pressed BN is a transversely isotropic material with $5$ independent elastic constants but can be approximated as an isotropic material with $3$ independent elastic constants. The Young’s modulus anisotropy is moderate (35\%) and an average value of 65 GPa is assumed, while the Poisson’s ratio is given by $0.21$. 

The tile is assumed to be fixed in displacement and rotation along its bottom and side surfaces, which are also considered thermally insulated. In contrast, the top surfaces are free to move, and two cooling mechanisms are imposed as thermal boundary conditions: (i) a radiative heat flux characterized by an emissivity of $\epsilon = 0.85$\,\cite{Adachi_1999}, and (ii) an evaporative cooling flux, with the associated mass flux computed with the Knudsen formula\,\cite{Knudsen_1950}. It is emphasized that the vaporization properties of hot-pressed BN remain somewhat unresolved in the literature\,\cite{Dreger_1962,Hildenbrand_1963,Rovner_1976}. It is not clear whether BN has a sublimation temperature (ejection of gaseous BN) or dissociation temperature (ejection of N$_2$ molecules leaving a B remnant) around 3300 K. In a similar fashion, at lower temperatures, it is not clear whether BN vaporizes via sublimation or dissociation, which impacts the enthalpy of vaporization and the vapor pressure dependence on the temperature. For simplicity, the Antoine correlation for the vapor pressure of B is adopted\,\cite{Yaws_2015} as well as the boron enthalpy of vaporization\,\cite{Chase_1998}.

To evaluate the onset of brittle failure, the Rankine or Coulomb criterion \cite{Budynas_1999} is employed. In this local criterion, the principal stresses $\sigma_{1,2,3}$ (stress tensor eigenvalues), are compared with the ultimate tensile strength (UTS) if positive or the ultimate compressive strength (UCS) if negative.  For $\sigma_{i}>0$, the failure occurs when any $\sigma_{i}$ exceeds the UTS, i.e., $\sigma_{i}-\mathrm{UTS}>0$. For $\sigma_{i}<0$, the failure occurs when the magnitude of any $\sigma_{i}$ exceeds the UCS, i.e., $|\sigma_{i}|-\mathrm{UCS}>0$. The anisotropies in the ultimate tensile strength (30\%) and ultimate compressive strength (25\%) are moderate and the average values of $\mathrm{UTS}=75\,$MPa and $\mathrm{UCS}=172\,$MPa are employed\,\cite{Pierson_1996}. Due to the lack of reliable $\mathrm{UTS}$ and $\mathrm{UCS}$ data above the room temperature, no temperature dependence is assumed for the Rankine criterion, unlike in the modeling of graphite\,\cite{Ratynskaia_2025}.

\section{Results} \label{ssec:results}

As aforementioned, the lack of information concerning the incident REs necessitates a parametric scan of the relevant impact characteristics, namely the initial kinetic energy, pitch angle, and tile wetted area. Here we choose a representative low-energy value of 2\,MeV and representative high-energy values of 10\,MeV and 20\,MeV. The initial pitch angle plays a crucial role in determining the effective angle of incidence on the tile and is therefore scanned from 0$^\circ$ up to 25$^\circ$. As far as the wetting is concerned, two different case studies are considered: loading only the flat top surface and loading the flat top along with the side surface over approximately half its height. 

The final result of the modeling work-flow, the volume of brittle failure, is compared with the empirical evidence in terms of the following two aspects; (i) the symmetry with respect to the main axis of the tile, (ii) the cumulative damage, that is a result of several discharges, which extends to a maximum depth of $\sim$1\,mm. The latter implies a failure depth of $\sim 100$ $\mu$m
per discharge. Note that here we model the onset of failure, hence the simulations are stopped right after the failure criterion is met\,\cite{Ratynskaia_2025}. However, in cases when the criterion is met early, it is expected that the material will continue to fail and release debris until the end of the loading ($1\,$ms), see Sec.\ref{sec:discussion} for more details.

\subsection{Energy deposition and temperature profiles}

\begin{figure}[!h]
    %\centering
    \raggedright
    \addtolength{\tabcolsep}{-9pt} 
    \begin{tabular}{c}
        \subfloat{%
          \begin{overpic}[width = 3.5in]{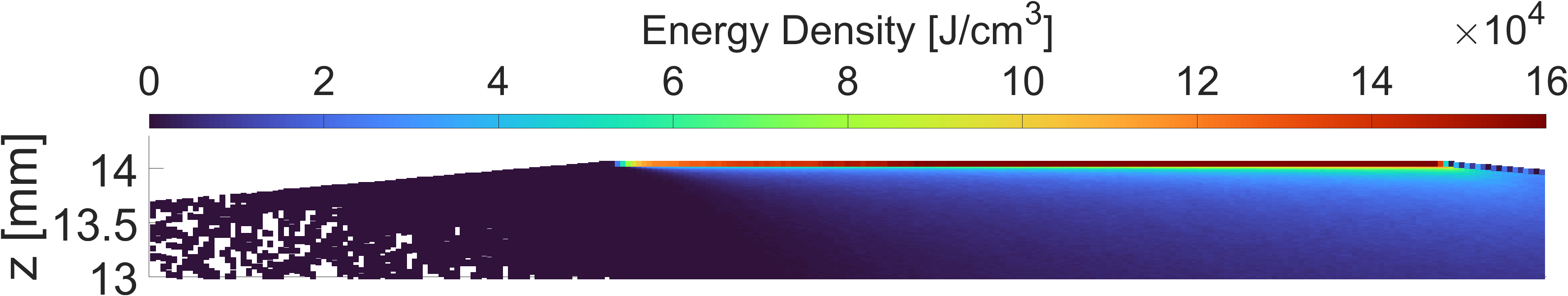}
          \put(10,15){\hbox{\kern3pt\textcolor{red}{\textbf{a)} }}}
          \end{overpic}
        }
        \\
        \subfloat{%
          \begin{overpic}[width = 3.5in]{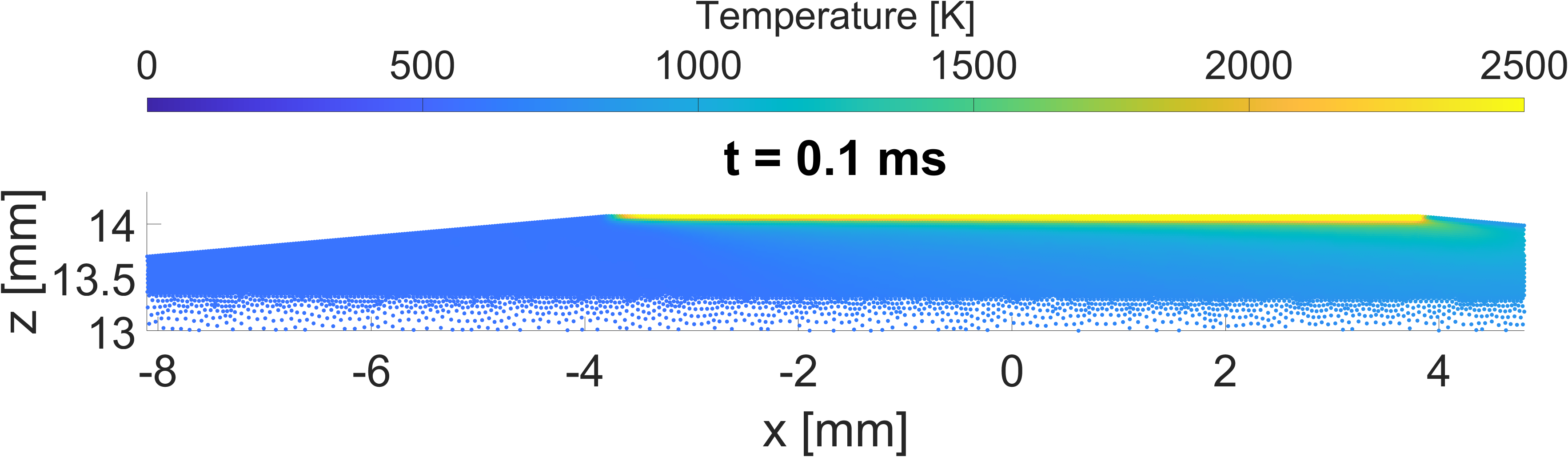}
          \put(10,15){\hbox{\kern3pt\textcolor{red}{\textbf{b)} }}}
          \end{overpic}
        }
        \\
    \end{tabular}
    \addtolength{\tabcolsep}{9pt} 
    \caption{Cross-section view. Simulation results for the energy density (a) and the temperature (b) for the case of $2\,$MeV \& 0$^\circ$ pitch angle.  The RE beam is coming from the left, only the top is loaded.}
    \label{fig:Temp_energy_2MeV}
\end{figure}

\begin{figure}[!h]
    %\centering
    \raggedright
    \addtolength{\tabcolsep}{-9pt} 
    \begin{tabular}{c}
        \subfloat{%
          \begin{overpic}[width = 3.5in]{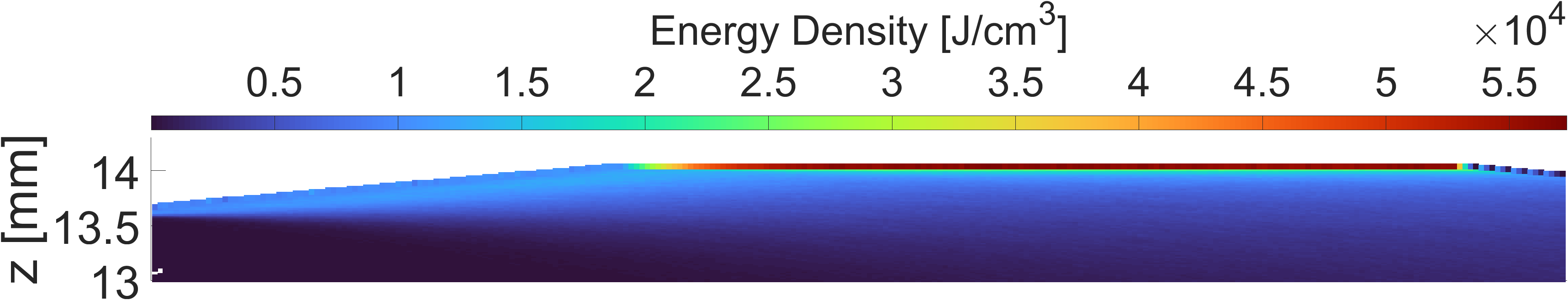}
          \put(10,15){\hbox{\kern3pt\textcolor{red}{\textbf{a)} }}}
          \end{overpic}
        }
        \\
        \subfloat{%
          \begin{overpic}[width = 3.5in]{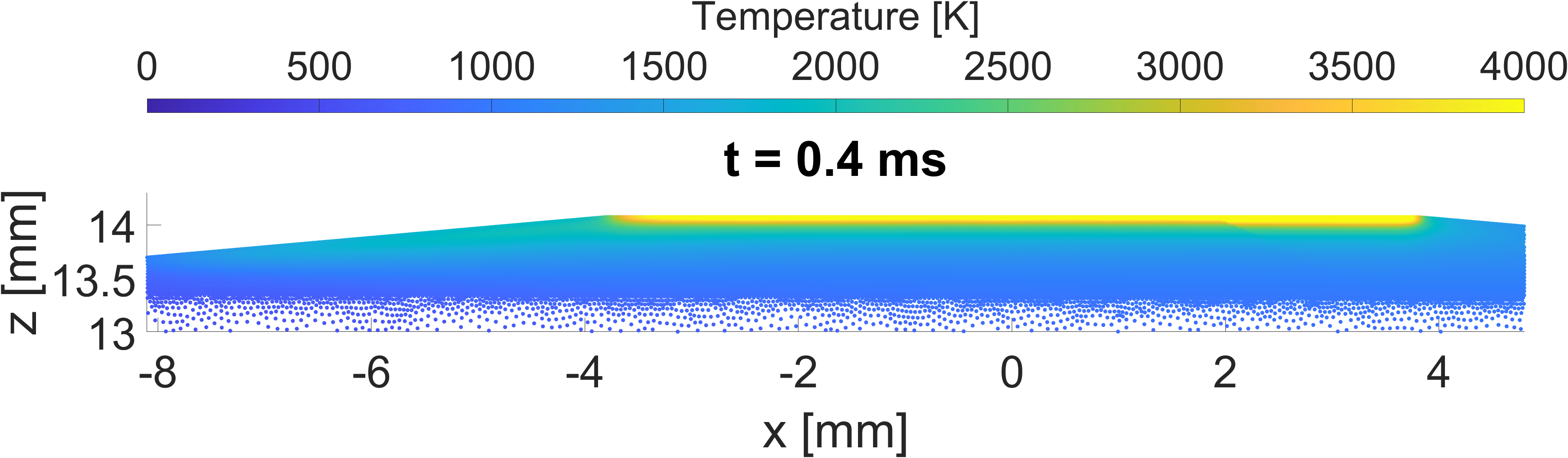}
          \put(10,15){\hbox{\kern3pt\textcolor{red}{\textbf{b)} }}}
          \end{overpic}
        }
        \\
    \end{tabular}
    \addtolength{\tabcolsep}{9pt} 
    \caption{Cross-section view. Simulation results for the energy density (a) and the temperature (b) for the case of $20\,$MeV \& 0$^\circ$ pitch angle. The RE beam is coming from the left, side and top are loaded.}
    \label{fig:Temp_energy_20MeV}
\end{figure}

The simulation results for the energy deposition and temperature are shown in Fig.\ref{fig:Temp_energy_2MeV} and Fig.\ref{fig:Temp_energy_20MeV}, for the two representative scenarios of 2 MeV and 20 MeV initial kinetic energy and zero pitch angle. Due to the very shallow angle of incidence, the energy deposition is limited in both cases to the first few 100's of $\mu$m, despite the much larger CSDA range values of $0.57\,$cm and $5.6\,$cm, respectively. Given the short loading time and limited effect of thermal diffusion, the temperature maps essentially reflect the energy deposition profiles, with the steepest gradients in the near-surface layers. As we shall see in what follows, thermal-expansion driven brittle failure will also be constrained to a similar volume. 

We recall that here, to facilitate the comparison between the cases studied, all energy density maps are normalized to $20\,$kJ deposited across the entire tile, i.e. over the full length of $132\,$mm. In general, the incident energy is larger than the absorbed energy due to the escaping primary particles (electrons) and/or secondary particles (mainly photons but also delta electrons or positrons). In particular, the electron backscattering yields are significantly enhanced at the grazing angles\,\cite{Bronshtein_1969,Tolias_2014} and the fraction of backscattered electrons returning to the surface depends on the tile and magnetic field geometry\,\cite{Ratynskaia_2025_R,Bartels1994,Kunugi_1993}. In the scenarios presented here, we find that for the low-energy RE case of Fig.\ref{fig:Temp_energy_2MeV}, the total energy losses are $\sim20\%$, out of which $\sim$17\% are due to electron backscattering and $\sim$3\% due to photon emission. In the 20 MeV case, depicted in Fig.\ref{fig:Temp_energy_20MeV}, the energy losses are much higher ($\sim60\%$) again primarily due to electron backscattering ($\sim$51\%) with a small contribution from photon losses ($\sim$9\%). 

\subsection{Failure due to low energy runaway electrons}

The simulations of the thermo-mechanical response due to impacts by the low-energy part of the RE energy distribution are carried out with REs of $2\,$MeV. When only the top flat surface of the tile is impacted, the predicted failure region is nearly symmetric, for both initial pitch angles of 0$^\circ$ and 10$^\circ$, as shown in Fig.\ref{fig:matching_scenario}(a). The corresponding failure depths are approximately $\sim$30\,$\mu$m and $\sim$60\,$\mu$m, respectively.

In contrast, when the tile edge is also exposed to the RE beam, the initial pitch angle plays a more critical role. A pitch angle of 0$^\circ$ results in shallow impact trajectories, leading to localized energy deposition with high energy density. This produces highly localized elevated temperatures and stresses, causing failure in the edge region within a layer of $\sim$100\,$\mu$m thickness, as shown in Fig.\ref{fig:non_matching}(a). A similar behavior is observed for an initial pitch angle of 10$^\circ$, where failure at the edge is predicted to occur only $0.2\,$ms later than at the top surface. Conversely, with a pitch angle of 25$^\circ$, the REs strike the surface at a steeper angle, leading to a more relaxed energy deposition. In this case, no failure is predicted at the tile edge, while a thicker failure layer of $\sim$100\,$\mu$m is expected on the top surface, as illustrated in Fig.\ref{fig:matching_scenario}(b).

Overall, among the scenarios tested, low energy RE beams lead to a region of failure, which is consistent with the experiment for all initial pitch angles when wetting occurs exclusively on the tile flat top, and for initial pitch angles greater than $\sim$25$^\circ$ when the side of the tile is also loaded.

\subsection{Failure due to high energy runaway electrons}

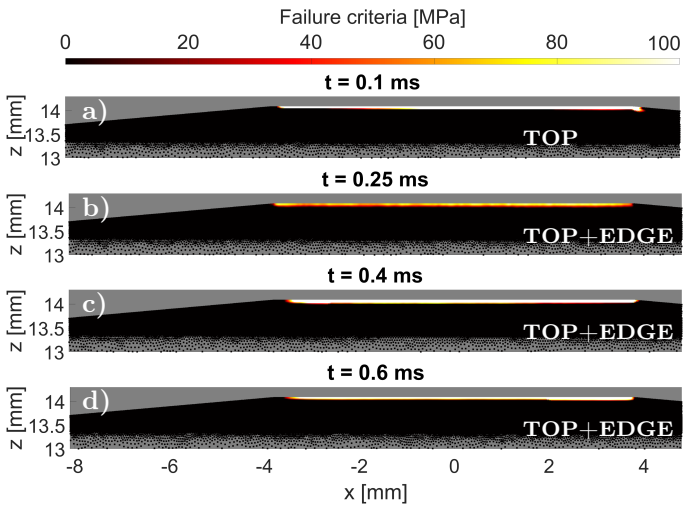
\begin{figure}[!h]
    %\centering
    \raggedright
    \addtolength{\tabcolsep}{-9pt} 
    \begin{tabular}{c}
        \subfloat{%
          \begin{overpic}[width = 3.5in]{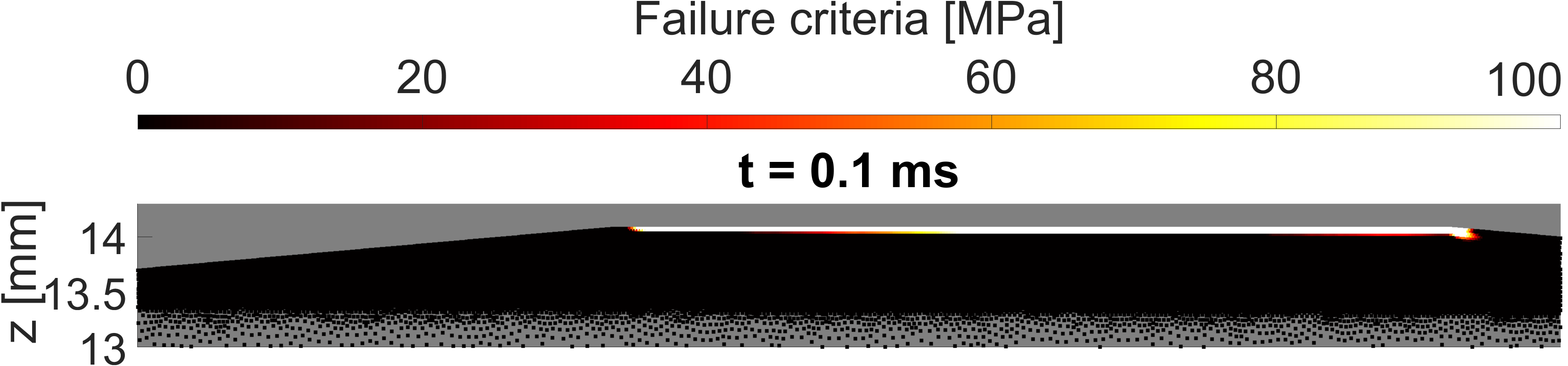}
          \put(10,7){\hbox{\kern3pt\textcolor{white}{\textbf{a)} }}}
          \put(76,3.1){\hbox{\kern1pt\textcolor{white}{\textbf{\footnotesize {TOP}} }}}
          \end{overpic}
        }
        \\
        \subfloat{%
          \begin{overpic}[width = 3.5in]{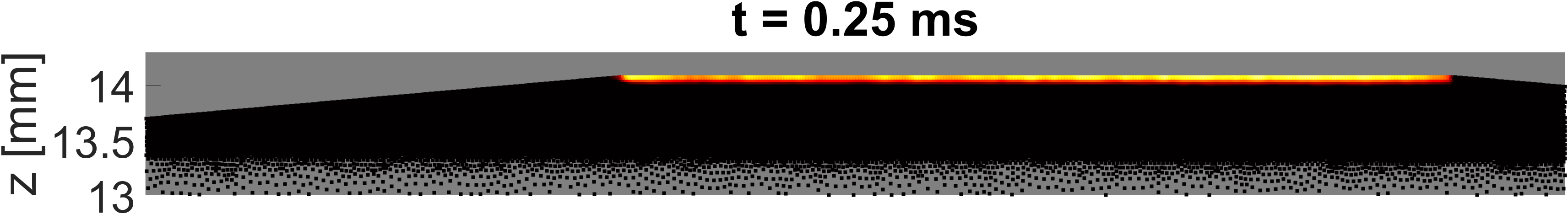}
          \put(10,7){\hbox{\kern3pt\textcolor{white}{\textbf{b)} }}}
          \put(76,2.9){\hbox{\kern1pt\textcolor{white}{\textbf{\footnotesize {TOP+EDGE}} }}}
          \end{overpic}
        }
        \\
        \subfloat{%
          \begin{overpic}[width = 3.5in]{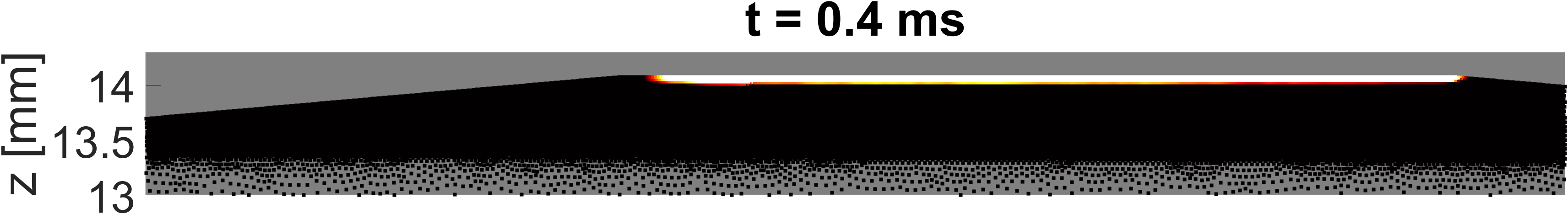}
          \put(10,7){\hbox{\kern3pt\textcolor{white}{\textbf{c)} }}}
          \put(76,3.1){\hbox{\kern1pt\textcolor{white}{\textbf{\footnotesize {TOP+EDGE}} }}}
          \end{overpic}
        }
        \\
        \subfloat{%
          \begin{overpic}[width = 3.5in]{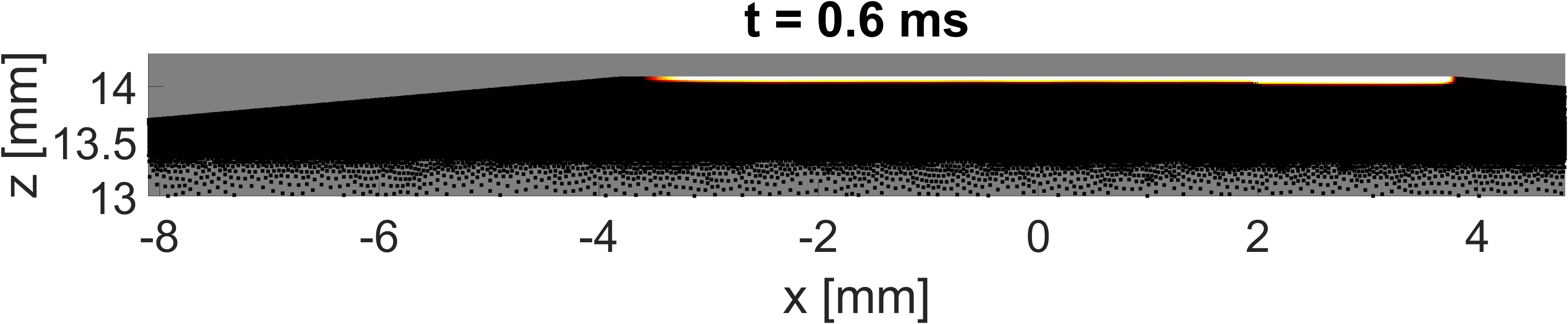}
          \put(10,14){\hbox{\kern3pt\textcolor{white}{\textbf{d)} }}}
          \put(76,10.0){\hbox{\kern1pt\textcolor{white}{\textbf{\footnotesize {TOP+EDGE}} }}}
          \end{overpic}
        }\\
    \end{tabular}
    \addtolength{\tabcolsep}{9pt} 
    \caption{Cross-section view. Modeling results for region of failure, with loading specifications as indicated in the images. Cases of  $2\,$MeV \& 0$^\circ$ pitch (a), $2\,$MeV \& 25$^\circ$ pitch (b), $20\,$MeV \& 0$^\circ$ pitch (c), $10\,$MeV \& 10$^\circ$ pitch (d). Indicated is also the instant of failure: $t=0.1\,$ms (a), $t=0.25\,$ms (b), $t=0.4\,$ms (c), $t=0.6\,$ms (d). The RE beam is coming from the left.}
    \label{fig:matching_scenario}
\end{figure}

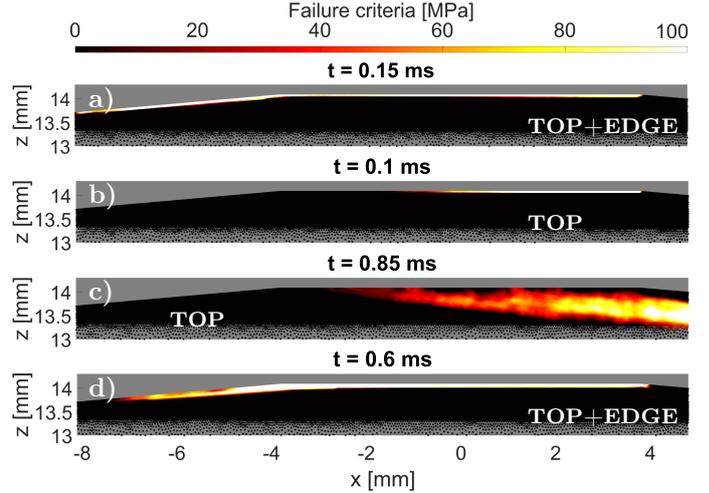
\begin{figure}[!h]
    %\centering
    \raggedright
    \addtolength{\tabcolsep}{-9pt} 
    \begin{tabular}{c}
        \subfloat{%
          \begin{overpic}[width = 3.5in]{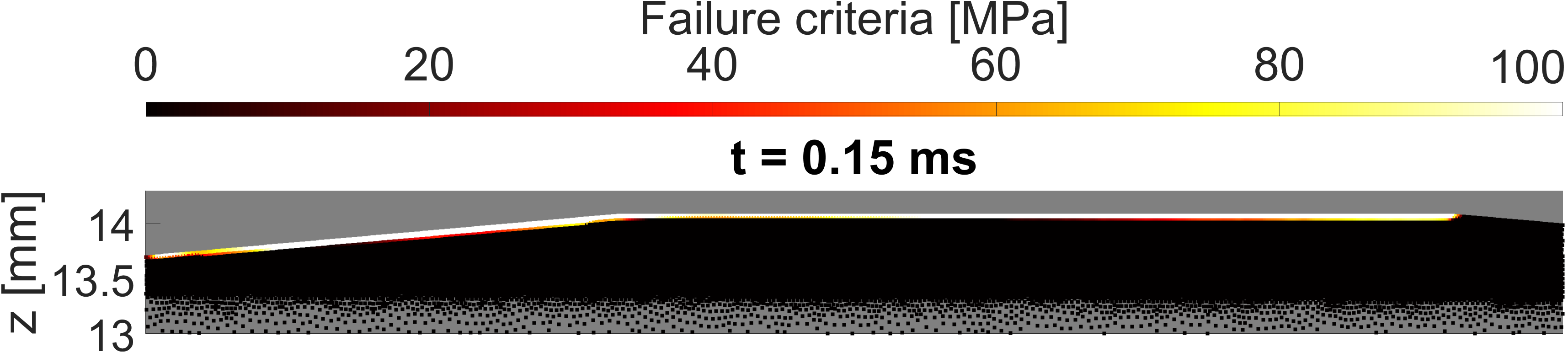}
          \put(10,7){\hbox{\kern3pt\textcolor{white}{\textbf{a)} }}}
          \put(75,3.1){\hbox{\kern3pt\textcolor{white}{\textbf{\footnotesize {TOP+EDGE}} }}}
          \end{overpic}
        }
        \\
        \subfloat{%
          \begin{overpic}[width = 3.5in]{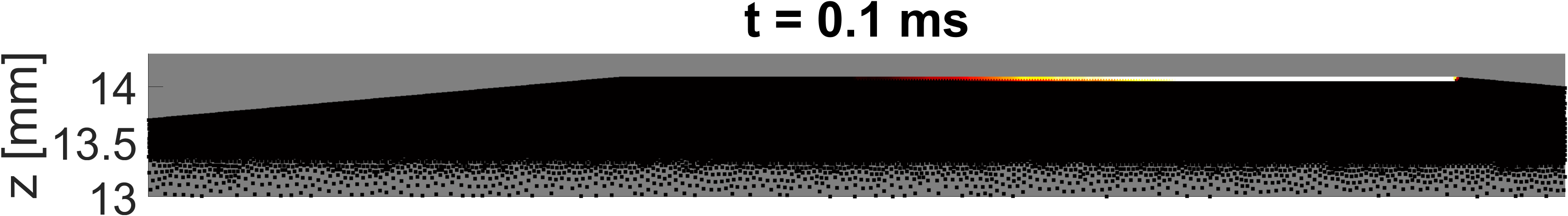}
          \put(10,7){\hbox{\kern3pt\textcolor{white}{\textbf{b)} }}}
          \put(75,3.1){\hbox{\kern3pt\textcolor{white}{\textbf{\footnotesize {TOP}} }}}
          \end{overpic}
        }
        \\
        \subfloat{%
          \begin{overpic}[width = 3.5in]{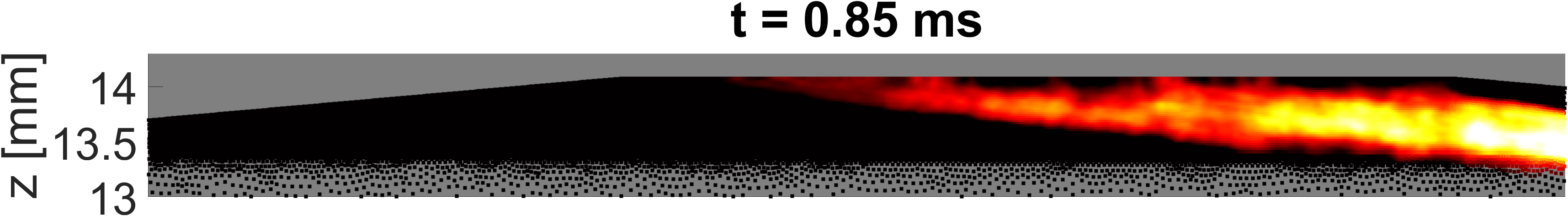}
          \put(10,7){\hbox{\kern3pt\textcolor{white}{\textbf{c)} }}}
          \put(22,3.1){\hbox{\kern3pt\textcolor{white}{\textbf{\footnotesize {TOP}} }}}
          \end{overpic}
        }
        \\
        \subfloat{%
          \begin{overpic}[width = 3.5in]{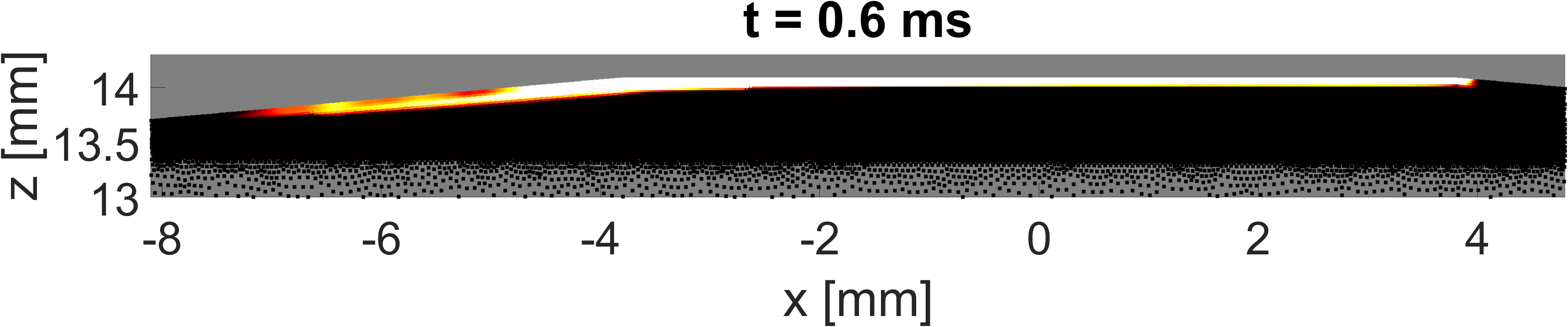}
          \put(10,14){\hbox{\kern3pt\textcolor{white}{\textbf{d)} }}}
          \put(75,10.1){\hbox{\kern3pt\textcolor{white}{\textbf{\footnotesize {TOP+EDGE}} }}}
          \end{overpic}
        }\\
    \end{tabular}
    \addtolength{\tabcolsep}{9pt} 
    \caption{Cross-section view. Modeling results for region of failure, with loading specifications as indicated in the images.  Cases of $2\,$MeV \& 0$^\circ$ pitch (a), $20\,$MeV \& 0$^\circ$ pitch (b), $20\, $MeV \& 10$^\circ$ impact (c), $10\,$MeV \& 0$^\circ$ pitch (d). Indicated is also the instant of failure: $t=0.15\,$ms (a), $t=0.1\,$ms (b), $t=0.85\,$ms (c),  $t=0.6\,$ms (d). The RE beam is coming from the left.}
    \label{fig:non_matching}
\end{figure}

The simulations of the thermo-mechanical response due to impacts by the high-energy part of the RE energy distribution are carried out with REs of 10\,MeV and 20\,MeV. At these energies, the electron CSDA range in BN increases significantly, resulting in lower, with respect to the 2 MeV case, energy densities and temperatures. Consequently, a simulation with 20\,MeV initial energy and 0$^\circ$ pitch angle predicts no failure on the tile edge, even when that surface is directly exposed to the RE beam. This is illustrated in Fig.\ref{fig:matching_scenario} (c), where the failed region has a thickness of approximately 70\,$\mu$m. However, when the RE loading is restricted to the tile top surface, the resulting failure region lacks symmetry with respect to the main axis, as shown in Fig.\ref{fig:non_matching}(b), and is therefore inconsistent with the observations.

A specific case is also analyzed for an initial energy of 20\,MeV with an imposed impact angle of 10$^\circ$. Under these conditions, the steep incidence angle causes the energy deposition to extend deeper into the tile. The resulting energy density is low, and failure occurs near the end of the loading period, nearly 1\,mm below the surface. The failure region is also clearly asymmetric, as depicted in Fig.~\ref{fig:non_matching}(c). This case highlights the importance of the shallow impact angle associated with the 0.3$^\circ$ B-field inclination, which better matches the observed damage morphology.

Further sensitivity of the model to the RE impact parameters is demonstrated through simulations with an initial kinetic energy of 10\,MeV. Two scenarios are considered, corresponding to initial pitch angles of 0$^\circ$ and 10$^\circ$, with the RE beam also impacting the tile edge.
In the case of a 10$^\circ$ pitch angle, the predicted failure is symmetric and confined to the top surface of the tile, with a thickness of approximately 80\,$\mu$m, as shown in Fig.\ref{fig:matching_scenario} (d). Conversely, a pitch angle of 0$^\circ$ leads to an initially confined failure at the top surface, which subsequently propagates to the tile edge after 0.3\,ms, as illustrated in Fig.\ref{fig:non_matching} (d).

Overall, from the scenarios tested, it can be concluded that, in order to reproduce observed symmetric damage, high energy RE beam requires loading the edge of the tile, in addition to the flat top. However, in case of more modest energies, e.g. 10 MeV, also non zero ($\geq$10$^\circ$) pitch angles  are required to ensure symmetric failure.

%for loading as indicated in the images

\subsection{Consistency with IR camera measurements of surface temperature evolution}

The temporal evolution of the surface temperature, reconstructed from the wide-angle IR camera measurements, is employed here for comparison with the outcome of the COMSOL simulations. The measurements concern three distinct discharges (\#60179, \#60180, \#60182), all exhibiting RE terminations on sector PJ5, tile B6 of the WEST inner limiter. The reconstructed surface temperature is averaged over the entire surface of tile B6.

The uncertainty in the emissivity of boron nitride at high temperatures, which can have values between 0.85 and 1.0, introduces a small, less than 5\%, error in the temperature,  depicted as a shadow over the solid-line curves in Figs.~\ref{fig:IR_temperature_noLoad}-\ref{fig:IR_temperature_Load}. However, the camera saturates at temperatures above $\sim2400$\,K, beyond which the recorded signal is unreliable. Hence, IR measurements are used primarily to analyze the seconds-long decay of the surface temperature.

To enable comparison with these measurements, long-time scale COMSOL simulations are performed, incorporating \emph{ad hoc} assumptions to mimic different scenarios of the material failure and loading.

\subsubsection{Loading termination after material failure}

A first set of simulations is performed using the Geant4 energy deposition map corresponding to the scenario with 20\,MeV initial kinetic energy, 0$^\circ$ pitch angle, and wetting also on the tile side. As discussed in the previous section, this configuration well matches with the observations.

\begin{figure}[!h]
    \centering
    \includegraphics[width=0.8\linewidth]{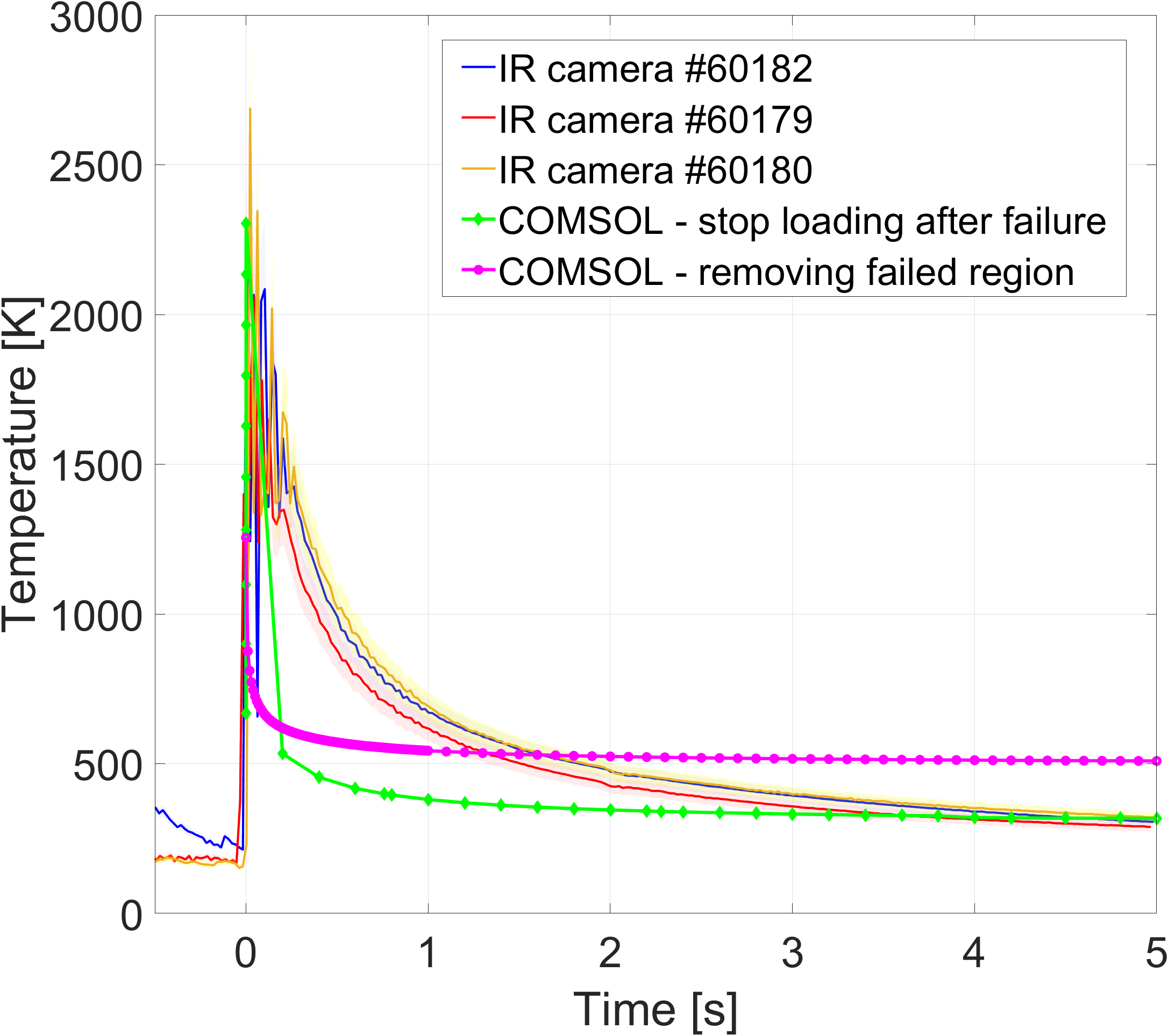}
    \caption{IR camera measurements of the tile surface temperature for three different pulses. Modeling results for RE impact parameters of 20 MeV \& 0$^\circ$ pitch. In the simulations. loading is terminated after failure and the temperature is monitored on the pristine surface (green curve) or the newly exposed surface after removing the failed region (pink curve).}
    \label{fig:IR_temperature_noLoad}
\end{figure}

In the first COMSOL simulation, the loading is applied only up to the time of the failure onset (i.e.\, 0.4\,ms), and the temperature is monitored at the top surface,  located centrally on the tile. Due to the shallow energy deposition in this scenario, heat is rapidly conducted into the cooler underlying material, resulting in a fast temperature decay, as illustrated in Fig.~\ref{fig:IR_temperature_noLoad} by the green curve.

A second simulation attempts to replicate the effect of debris ejection by removing the region expected to fail at 0.4\,ms. The simulation is then continued on the modified (truncated) domain, and the temperature is monitored at the top of the newly exposed surface. In this case (Fig.~\ref{fig:IR_temperature_noLoad},the pink curve), however, the peak temperature reached is significantly lower than the measured one, since the most heavily loaded volume has been excluded from the domain. 

\subsubsection{Loading continuation after material failure}

A second set of simulations is performed without interrupting the loading at the onset of failure. Instead, the region predicted to fail according to the Rankine criterion is removed at the end of the anticipated tile-RE interaction, i.e., at 1\,ms. This approach serves as a crude approximation to account for the fact that debris ejection does not occur instantaneously as well as the failure region can continue to grow after the Rankine criterion is met initially. Moreover, it allows for the complete deposition of 20\,kJ into the tile.

\begin{figure}[!h]
    \centering
    \includegraphics[width=0.8\linewidth]{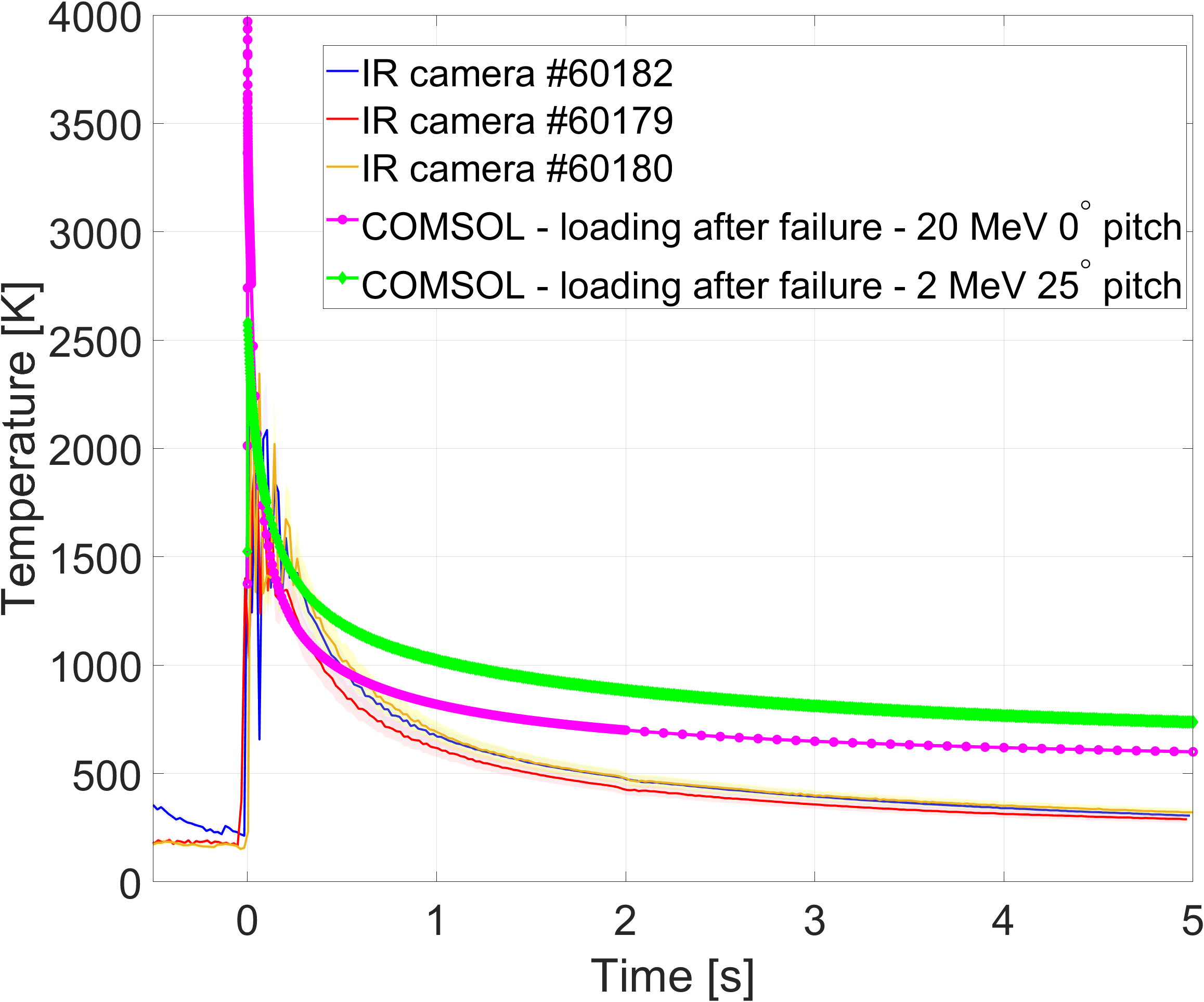}
    \caption{IR camera measurements of the tile surface temperature for three different pulses. Modeling results for two different sets of RE impact parameters (specified in the legend). In the simulations, loading continues (up to 1 ms) after the failed region is removed.}
    \label{fig:IR_temperature_Load}
\end{figure}

Two different energy deposition maps are considered: the previously discussed 20\,MeV, 0$^\circ$ pitch angle scenario and the 2\,MeV, 25$^\circ$ pitch angle case. In both cases, the resulting temperature decay is slower and matches reasonably well the IR camera data during the first seconds after the impact, see pink and green curves in Fig.\ref{fig:IR_temperature_Load}.  However, beyond $t \sim 1$\,s, the simulated temperature profile flattens, in contrast to the measurements. This discrepancy is attributed to the employed boundary conditions. Specifically, the tile is assumed to be thermally insulated on both side and bottom surfaces, whereas in reality it is in thermal contact with the surrounding structures, allowing for continuous heat dissipation.

\section{Discussion and conclusions}{\label{sec:discussion}}

Accidental RE-induced damage on the WEST inner limiter boron nitride tiles has been modeled employing a multi-physics workflow, recently developed and validated in the controlled DIII-D graphite dome experiment\,\cite{Ratynskaia_2025}. The approach couples Geant4-based Monte Carlo simulations of RE energy deposition with COMSOL-based finite-element analysis of the thermo-mechanical tile response, allowing for predictions of the onset of brittle failure based on the Rankine criterion.

Due to the accidental nature of the damage and the lack of direct measurements, the loading and impact parameters had to be deduced from available empirical data. In particular, since details of the energy distribution of the incoming REs are not available, the initial RE energies were scanned, including the high $\geq8$\,MeV part of the RE distribution known to be generated during WEST disruptions. The simulations have shown that, with reasonable pitch angle and wetting area variations, both low and high energy REs can inflict thermo-mechanical response consistent with the observations in terms of the damage topology and the material loss volume. Since Monte Carlo simulations for different RE energies are independent, in principle, the contributions of low and high energy electrons could be combined to simulate more complicated scenarios but the unknown ratio between the number of low and high energy electrons makes it an underdetermined problem.

The lack of empirical constraints also prevents the modeling of the non-linear stage of the interaction, where large deformations lead to fragmentation and debris ejection, as very recently done in simulations of the RE-induced explosion of the graphite dome in the controlled DIII-D experiment\,\cite{ITPA_Rizzi, EPS_Ratynskaia}. These simulations revealed that after the initial onset of failure, the fragmentation region grows and debris ejection is continuous up to the end of loading. This implies that the thickness of the initially failed regions found in the linear thermo-elastic modeling of brittle BN reported here, in the range of 60-120 $\mu$m, matches well the observations of $1\,$mm deep damage which is a cumulative loss over several discharges. This is also supported by the simulations of the long-time decay of the tile surface temperature, which agree with the IR camera measurements only when the energy loading is maintained throughout the full $1$ms of the RE-tile interaction, followed by the removal of the failed material volume. This comparison also serves as an additional consistency test for the chosen total deposited energy value.  

%While the approaches adopted in this work  offers a reasonable first-order approximation, it remains a crude representation of the actual situation. Capturing the dynamics of material fragmentation and debris ejection with higher fidelity requires more advanced simulations incorporating detailed failure and post-failure physics, with suitable approaches for high deformation regimes.

The limitations of simulations of accidental PFC damage caused by the lack of empirical constraints also highlight the need for more accurate and detailed input data in future modeling efforts. In fact, the critical role of well-informed input on the RE impact parameters has been recognized and the first controlled experiment on RE-induced damage in bulk tungsten tiles was performed in WEST in April 2025 (C11 experimental campaign). The experiment featured several diagnostics to provide detailed characterization of the RE impact and loading; fast visible and IR cameras to infer the impact duration and debris dynamics, spectroscopic measurements to reconstruct the RE beam energy and pitch angle distributions as well as embedded thermocouples to estimate the deposited energy. The work-flow developed for RE-induced damage on brittle sublimating materials, first validated in a controlled DIII-D experiment with graphite\,\cite{Ratynskaia_2025, ITPA_Rizzi, EPS_Ratynskaia} and applied here to accidental WEST explosions on boron nitride, is currently being extended to include key features of the thermo-mechanical and hydrodynamic response of tungsten; a multiphase equation of state and advanced constitutive models that capture visco-plastic behavior\,\cite{Ratynskaia_2025_R}.

\section*{Acknowledgments}

\noindent  The work has been performed within the framework of the EUROfusion Consortium,\,funded by the European Union via the Euratom Research and Training Programme (Grant Agreement No\,101052200 - EUROfusion). The views and opinions expressed are however those of the authors only and do not necessarily reflect those of the European Union or the European Commission. The European Union or European Commission cannot be held responsible for them. The simulations were enabled by resources provided by the National Academic Infrastructure for Supercomputing in Sweden (NAISS) and the Swedish National Infrastructure for Computing (SNIC) at the NSC (Link\"oping University) partially funded by the Swedish Research Council through grant agreements No\,2022-06725 and No\,2018-05973.

\bibliography{biblio_new}

\end{document}